# Title: Electrical Control of Second-Harmonic Generation in a WSe$_2$ Monolayer Transistor

**Authors:** Kyle L. Seyler[1], John R. Schaibley[1], Pu Gong[2], Pasqual Rivera[1], Aaron M. Jones[1], Sanfeng Wu[1], Jiaqiang Yan[3,4], David G. Mandrus[3,4,5], Wang Yao[2], Xiaodong Xu[1,6*]

**Affiliations:**

[1]Department of Physics, University of Washington, Seattle, Washington 98195, USA
[2]Department of Physics and Center of Theoretical and Computational Physics, University of Hong Kong, Hong Kong, China
[3]Materials Science and Technology Division, Oak Ridge National Laboratory, Oak Ridge, Tennessee, 37831, USA
[4]Department of Materials Science and Engineering, University of Tennessee, Knoxville, Tennessee, 37996, USA
[5]Department of Physics and Astronomy, University of Tennessee, Knoxville, Tennessee 37996, USA
[6]Department of Materials Science and Engineering, University of Washington, Seattle, Washington, 98195, USA

*Correspondence to: xuxd@uw.edu

## Abstract

**Nonlinear optical frequency conversion, in which optical fields interact with a nonlinear medium to produce new field frequencies[1], is ubiquitous in modern photonic systems. However, the nonlinear electric susceptibilities that give rise to such phenomena are often challenging to tune in a given material, and so far, dynamical control of optical nonlinearities remains confined to research labs as a spectroscopic tool[2]. Here, we report a mechanism to electrically control second-order optical nonlinearities in monolayer WSe$_2$, an atomically thin semiconductor. We show that the intensity of second-harmonic generation at the A-exciton resonance is tunable by over an order of magnitude at low temperature and nearly a factor of 4 at room temperature through electrostatic doping in a field-effect transistor. Such tunability arises from the strong exciton charging effects in monolayer semiconductors[3,4], which allow for exceptional control over the oscillator strengths at the exciton and trion resonances. The exciton-enhanced second-harmonic generation is counter-circularly polarized to the excitation laser, arising from the combination of the two-photon and one-photon valley selection rules[5-8] that have opposite helicity in the monolayer. Our study paves the way towards a new platform for chip-scale, electrically tunable nonlinear optical devices based on two-dimensional semiconductors.**

**Main Text**

Electro-optic modulators are widely used in photonic devices for optical switching and communications. Linear optical properties, such as the refractive index and linear absorption, are controllable by a diversity of phenomena; this enables electro-optic modulators to find use from table-top laser systems to chip-scale optical interconnects. By extension, one may imagine similar schemes in which the nonlinear optical properties are controllable. The possibility of tuning the second-order electric susceptibility, $\chi^{(2)}$, is particularly appealing, as it is responsible for second-harmonic generation (SHG) and optical parametric amplification. An SHG transistor, where strong SHG signals can be electrically switched on and off by a gate, would extend the versatility of modern photonic systems and may enable new approaches to optical signal processing.

Electrically tunable SHG is a well-established concept, which was first demonstrated in calcite under an applied electric field over 50 years ago[2]. Electric-field-induced SHG, enabled by an electric field that breaks a material's inversion symmetry, has proven to be an indispensable tool in a great variety of fields. It is generally a weak effect, although recently it was shown that plasmonics[9] and metamaterials[10] hold promise for enhancing it. Other methods to dynamically control SHG include current-induced SHG[11], charge-assisted SHG tuning in a nanostructured plasmonic device[12], and resonant Stark tuning of intersubband transitions in coupled quantum wells[13]. However, new methods are critical in the effort to engineer electrically controllable nanoscale SHG devices for on-chip optical communications and tunable light sources. In this Letter, we demonstrate a new approach based on electrical tuning of the exciton charging effects in two-dimensional (2D) semiconductors.

Atomically thin group-VIB transition metal dichalcogenides (TMDs), such as $WSe_2$, have stimulated great interest due to their outstanding optoelectronic properties, which involve strongly bound[14-20] and tunable[3,4] excitons that show optically initialized valley polarization[5-8] and coherence[21]. Recently, there has been a surge of attention on the nonlinear optical properties of 2D TMDs[16-19, 22-28]. Some notable findings include the strong SHG of odd-layered exfoliated TMDs[22-26], in which inversion symmetry is broken, and the linear polarization dependence of the SHG intensity[22-24,26-28], which reflects the underlying $D_{3h}$ crystal symmetry. In addition, multiple studies have extracted the exceptionally large exciton binding energies of $WS_2$ and $WSe_2$ with the aid of two-photon-induced photoluminescence (PL) excitation spectroscopy[16-19]. However, electrical control of the coherent nonlinear optical response of 2D valley excitons remains unexplored.

Here, we demonstrate the electric control of SHG intensity in a monolayer $WSe_2$ field-effect transistor (FET), where the tunable SHG arises from the resonant responses of neutral and charged excitons with electrostatically controlled charging effects, a mechanism fundamentally different than electric-field-induced SHG. Our devices consist of mechanically exfoliated $WSe_2$ crystals on 300 nm $SiO_2$ on $n^+$-doped Si, with V/Au (5/50 nm) contacts patterned by standard electron beam lithography techniques (Figs. 1a and b). Measurements are performed at normal incidence in reflection geometry with a setup optimized for polarization-resolved micro-PL and -SHG (see Methods).

We first study the SHG at room temperature. Under excitation at 0.83 eV (~1.5 μm), we observe strong emission of the A exciton at 1.66 eV (~750 nm) that scales quadratically with excitation power (Fig. 1c). Since the excitation is two-photon resonant with the exciton energy, SHG and PL induced by two-photon absorption (2-P PL) are degenerate (Fig. 1d). However, the laser-like line shape of the emission suggests that SHG dominates, and the polarization dependence allows us to further rule out the possibility of 2-P PL. Figure 1e shows the emission intensity parallel to the incident laser polarization as a function of crystal orientation. We find a characteristic six-fold pattern, which is expected for SHG in monolayer TMDs[22-24,26,28] and reflects the threefold rotational symmetry of the crystal (Fig. 1e). In contrast, if the emission were 2-P PL, the intensity of the co-linearly polarized emission would be independent of crystal orientation. By exciting the sample with photon energy 50 meV above the exciton resonance, we observe 2-P PL and find that its polarization is independent of crystal axis as expected (Fig. 1d and 1e). From the polarization dependence of the emission at 1.66 eV, the ratio between the strongest and weakest emission is over 40, which shows that 2-P PL, if present, is much weaker than the SHG.

At room temperature, when the excitation energy is scanned across the two-photon resonance at the A exciton, the on-resonance SHG is over 15 times stronger than off-resonance, corresponding to a second-order susceptibility contrast of ~4 (Fig. 2b). Clearly, the strong SHG signal is from resonant coupling to the A exciton. We estimate an effective volume second-order susceptibility of ~60 pm/V on resonance (equivalent to sheet second-order susceptibility ~0.04 $nm^2$/V, see Methods and Fig. 2b), comparable to reports of SHG in monolayer $MoS_2$ under higher energy excitation[22,23]. This corresponds to an approximate conversion efficiency[1,22] of $4 \times 10^{-10}$ under our experimental conditions (see Methods), which is an order of magnitude larger than what one would achieve if standard transparent nonlinear crystals[1] could be scaled to the same thickness.

The resonant SHG can be tuned by electrostatic doping. We observe a nearly fourfold reduction in the resonant SHG intensity when the DC gate voltage ($V_g$) is swept from -80 to 80 V (Fig. 2c-d). We do not include transport characteristics here, as reliable data requires excellent electrical contacts, which is an open problem in the field. However, the effect is reminiscent of the PL under electrostatic doping (Supplementary Fig. 1), in which the oscillator strength at the exciton resonance decreases with increased electron doping[29]. For excitation 30 meV above and below the exciton resonance, $V_g$ has no effect on the SHG (Fig. 2d), supporting the idea that the SHG tuning originates from the modulation of oscillator strength at the exciton resonance. To gain deeper insight into this effect, we study the SHG at low temperature, where the tunability is substantially increased.

Figure 3a shows a sequence of SHG intensity maps as the two-photon excitation energy is swept from the 0.85 to 0.875 eV at 30 K, with the corresponding excitation laser spectrum attached. In each map, for $V_g$ near -40 V, there is a peak around 1.74 eV. Toward higher positive $V_g$, the feature is suppressed and a new peak appears around 1.71 eV for the left two maps. Furthermore, for negative $V_g$, the SHG peak shifts to lower energy. These spectral features bear strong similarity to the gate-dependent PL for $WSe_2$ monolayers[21]. By comparing to the PL intensity plot in Figure 3b, we assign neutral exciton ($X^0$) for the peak at 1.74 eV for $V_g$ near -40 V, negatively charged excitons (negative trions, $X^-$) near 1.71 eV at small positive $V_g$, and positive trions ($X^+$) at lower energy for high negative $V_g$. The strong presence of $X^0$ at negative voltages shows the sample is intrinsically electron-doped. The broad spectrum of the pulsed laser (~25 meV full width at half maximum) allows us to excite both $X^0$ and the trions at once. In Fig. 3a, SHG over $X^0$ energy dominates, even when the excitation is not centred on $X^0$, due to the large oscillator strength at the exciton resonance relative to at the trion resonance[21].

The spectra in Figure 3c illustrate the tunability of the SHG by gate bias. In analogy to an FET where electrical response is controlled by gate voltage, here we demonstrate a monolayer SHG transistor. The electrostatic doping tunes the strength of the neutral and charged excitonic resonances, and thus the strength and frequencies of the SHG response (Fig. 3d). Specifically, $X^0$

SHG decreases by over an order of magnitude from -30 V to 70 V. In the approximately linear regime from -20 V to 0 V, the modulation is over 10% per volt, comparable to plasmonic electric-field-induced SHG devices[9]. For the trions, the SHG tunes opposite to the exciton signal, as expected. We note that we find similar curves under different excitation energies and devices (Supplementary Fig. 2 and 3). The strong tunability illustrates the important influence that the exciton species have on the second-order susceptibility.

Microscopically, the exciton or trion plays the role of a resonant intermediate state for the photon scattering process. When the sample is electrostatically doped, the eigenenergy spectrum of these intermediate states changes due to the Coulomb interaction. At zero doping, $X^0$ is the eigenstate of Coulomb interaction. At very large doping, an electron-hole pair will find an extra carrier in its close proximity, thus the trion becomes the eigenstate of the Coulomb interaction. At moderate doping, exciton and trion coexist. Therefore, with doping, the joint density of states (and hence the optical oscillator strength) shifts from the exciton resonance to the trion resonance, which results in the observed electrostatic tuning of the SHG efficiency at a given energy. We briefly note that the $X^-$ PL resonance shifts appreciably with gate voltage, likely due to the quantum-confined Stark effect[30]. Thus, it may also be possible to utilize the Stark effect to tune $X^-$ SHG intensity at a fixed frequency. However, from Figure 3, it is clear the main mechanism arises from tuning the oscillator strength at the exciton and trion resonance.

Finally, we remark on the intriguing two-photon optical selection rules of the exciton-enhanced SHG. Figure 4a depicts how two $\sigma-$ photons at the fundamental frequency ($\omega$) generate a single $\sigma+$ photon at the second-harmonic frequency ($2\omega$) with near unity polarization, and the opposite holds for $\sigma+$ excitation. This is the expected SHG selection rule from $D_{3h}$ crystal symmetry. The production of a counter-rotating second harmonic is a general feature of materials with both broken inversion and threefold rotational symmetry[31], where the lattice supplies the angular momentum mismatch of the absorbed and emitted photons. However, given the valley-contrasting physics of monolayer TMDs, it is interesting to examine the microscopic origin of the SHG selection rule. The one-photon emission process from the valleys follows as reported previously[5-8], with $\sigma+$ ($\sigma-$) emitted from $+K$ ($-K$) points in the Brillouin zone. For the two-photon interband transition, we find a flip in the selection rule: two $\sigma-$ ($\sigma+$) photons can be simultaneously absorbed at $+K$ ($-K$) (see Supplementary Section S2). Thus, we are lead to the expected counter-circular SHG selection rule (Fig. 4b), which illustrates how the microscopic valley-contrasting physics is intimately related to the crystal symmetries.

We have shown that monolayer TMD FETs represent a new class of electrically tunable nonlinear optical devices, with strong SHG around the A exciton energy, allowing for efficient optical frequency doubling. Improved device engineering, including better gating schemes and waveguide integration, will enable enhanced tunability. Furthermore, bandgap engineering through alloying[32-34] and heterostructures means that a wide range of wavelengths are potentially accessible, including SHG of the telecom wavelengths. These CMOS-compatible monolayer SHG transistors may enable new applications in optical signal processing, on-chip nonlinear optical sources, and integrated photonic circuits.

**Methods**

**SHG and PL measurements**

All optical measurements were performed in the reflection geometry with the emission collected by a spectrometer and Si CCD. The samples were measured under vacuum in a helium-flow cryostat (temperatures specified in main text). For SHG, an optical parametric amplifier (Coherent OPA 9800), pumped by an amplified mode-locked Ti:sapphire laser (RegA 9000), produced tunable ultrafast pulses from 0.5 eV to 1 eV with ~25 meV bandwidth at 250 kHz, which were focused to a spot size of ~ 2 μm by a 50x near-IR objective lens (Olympus). Unless otherwise specified, 40 μW of average power was used. The quadratic power dependence of SHG was verified to ensure no sample degradation. For circular polarization measurements, a half- and quarter-wave Fresnel rhomb were used to produce circularly polarized light from the laser. The back-collected signal was sent through the same Fresnel rhombs and reflected off a near-normal dichroic beam splitter, after which the polarization components were analysed with an achromatic half-wave plate and linear polarizer before entering the spectrometer. For linear polarization measurements, the linear polarization of the fundamental beam was rotated and the components of the emission polarization were analysed. For PL, 20 μW of power from a 655 nm (1.89 eV) diode laser was used to excite the sample.

**Measurement of second-order susceptibility**

SHG from monolayer $WSe_2$ exfoliated onto fused quartz was measured and compared to SHG from the front surface of a bulk crystal of z-cut α-quartz following Reference 22 and 23. The SHG was measured without an analyser in front of the detector and thus the intensity was independent of the incident laser polarization due to the symmetries of the WSe$_2$ and quartz[22]. For monolayer WSe$_2$, we thus determined the modulus of the only nonzero second-order sheet susceptibility element: $\chi_s^{(2)} \equiv \chi_{s,xxx}^{(2)} = -\chi_{s,xyy}^{(2)} = -\chi_{s,yyx}^{(2)} = -\chi_{s,yxy}^{(2)}$ (where $x$ is along the armchair direction)[22]. We used $\chi_{quartz}^{(2)} = 2d_{11} = 0.6$ pm/V for α-quartz[35] and took it as a constant in our wavelength range (1400 to 1600 nm). To obtain the effective volume second-order susceptibility, we divided the calculated second-order sheet susceptibility, $|\chi_s^{(2)}|$, by the monolayer WSe$_2$ thickness of 0.7 nm. Our estimation of the SHG conversion efficiency[1,22] assumed our standard excitation condition of 40 µW average power with 200 fs pulses at 250 kHz focused to a spot size of 2 µm, corresponding to a peak pulse intensity of ~24 GW/cm$^2$. These measurements were performed at room temperature.

**Acknowledgments:** This work is mainly supported by the DoE BES (DE-SC0008145 and SC0012509). Device fabrication is partially supported by NSF (DMR-1150719). P.G. and W.Y. were supported by the Croucher Foundation (Croucher Innovation Award), and the RGC and UGC of Hong Kong (HKU705513P, HKU9/CRF/13G, AoE/P-04/08). J.Y. and D.M. were supported by US DoE, BES, Materials Sciences and Engineering Division. X.X. thanks the support from Cottrell Scholar Award. S.W. thanks the support from State of Washington funded Clean Energy Institute.

Device fabrication was performed at the Washington Nanofabrication Facility and NSF-funded Nanotech User Facility.

**Author Contributions:** X.X. conceived the idea. K.L.S. designed the experiment and performed the measurements, assisted by J.R.S., A.M.J., and P.R.. P.R. and K.L.S. fabricated the devices. J.Y. and D.M. synthesized and characterized the bulk $WSe_2$ crystal. K.L.S. performed data analysis, with input from P.G., J.R.S., S.W., X.X., and W.Y.. K.L.S. wrote the paper, assisted by X.X., J.R.S., and W.Y.. All authors discussed the results and commented on the manuscript.

**Competing Financial Interests:** The authors declare no competing financial interests.

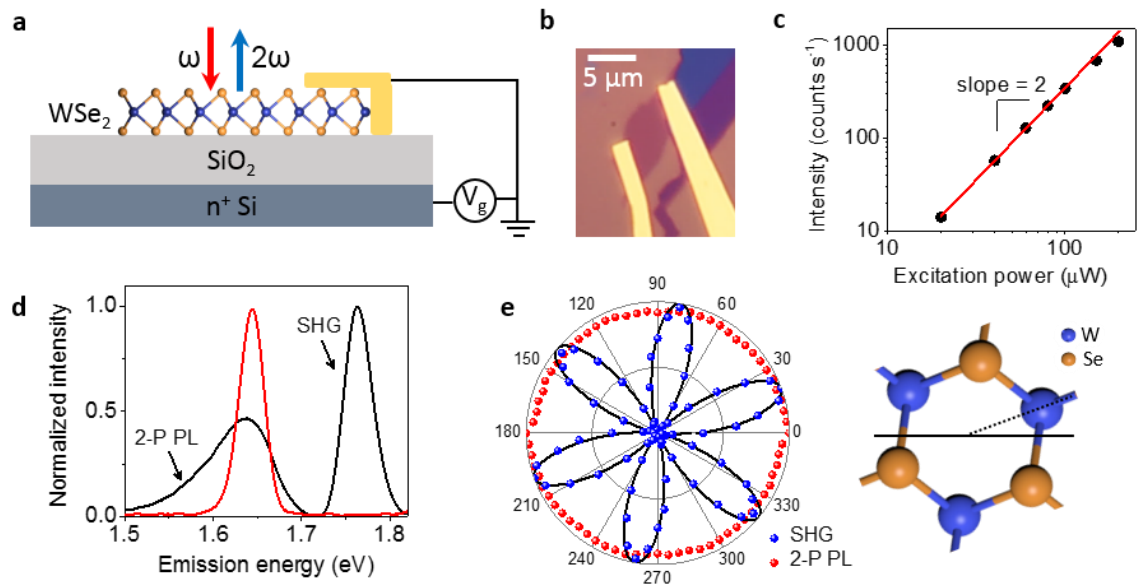

**Figure 1| Characterization of second-harmonic generation in monolayer WSe$_2$ transistor. a**, Schematic of gated monolayer WSe$_2$. Excitation at $\omega$ (red arrow) generates second-harmonic radiation at $2\omega$ (blue arrow). **b**, Microscope image of monolayer WSe$_2$ transistor. The light purple area between the contacts is monolayer WSe$_2$. **c**, Power dependence of SHG peak intensity for two-photon-resonant excitation of the A exciton, where the red line shows the expected quadratic dependence. **d**, Emission spectrum for excitation at 0.83 eV (red curve) dominated by SHG, and 0.88 eV (black curve), showing SHG at 1.76 eV and two-photon-induced PL (2-P PL) from the exciton. **e**, Resonant SHG and 2-P PL intensity (under non-resonant excitation at 0.88 eV) parallel to the incident laser polarization as a function of the crystal angle, with the black line showing the expected SHG polarization dependence. Top view of monolayer TMD crystal structure showing possible corresponding orientation.

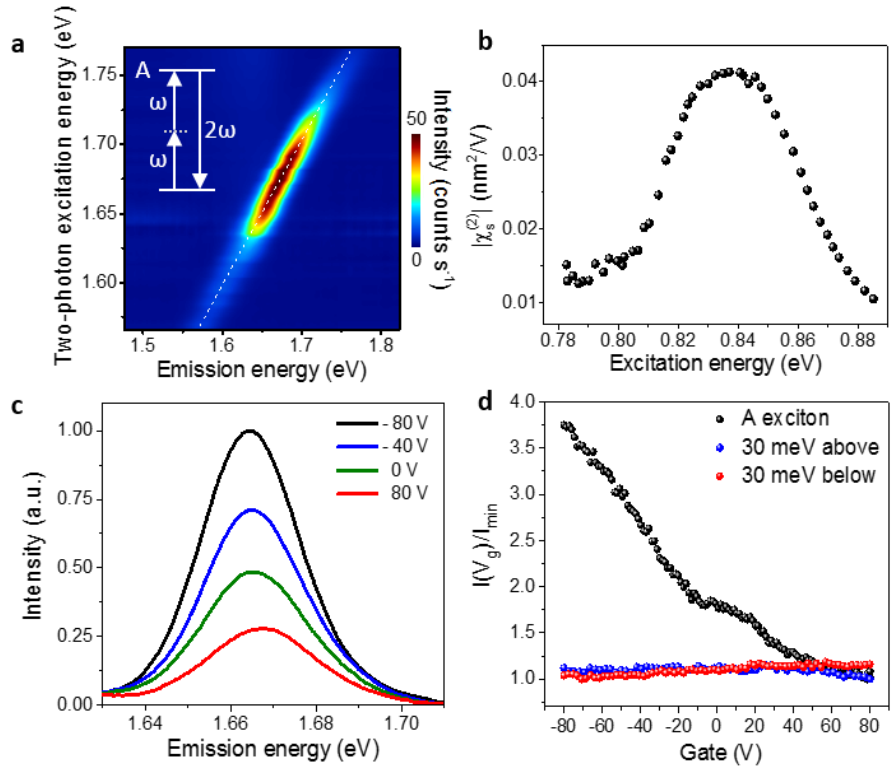

**Figure 2| Resonant enhancement and gate tunability of room-temperature SHG. a**, Two-photon excitation map showing the strong SHG resonance and negligible two-photon PL near the A exciton energy. Dashed line has a slope of unity. Inset: Energy level diagram of SHG resonant with the A exciton. **b**, Experimentally measured sheet second-order susceptibility as a function of excitation energy. **c**, SHG spectra on resonance with exciton at selected gate voltages. **d**, Normalized peak intensity of SHG as a function of gate voltage with two-photon excitation energy above, below, and resonant with A exciton.

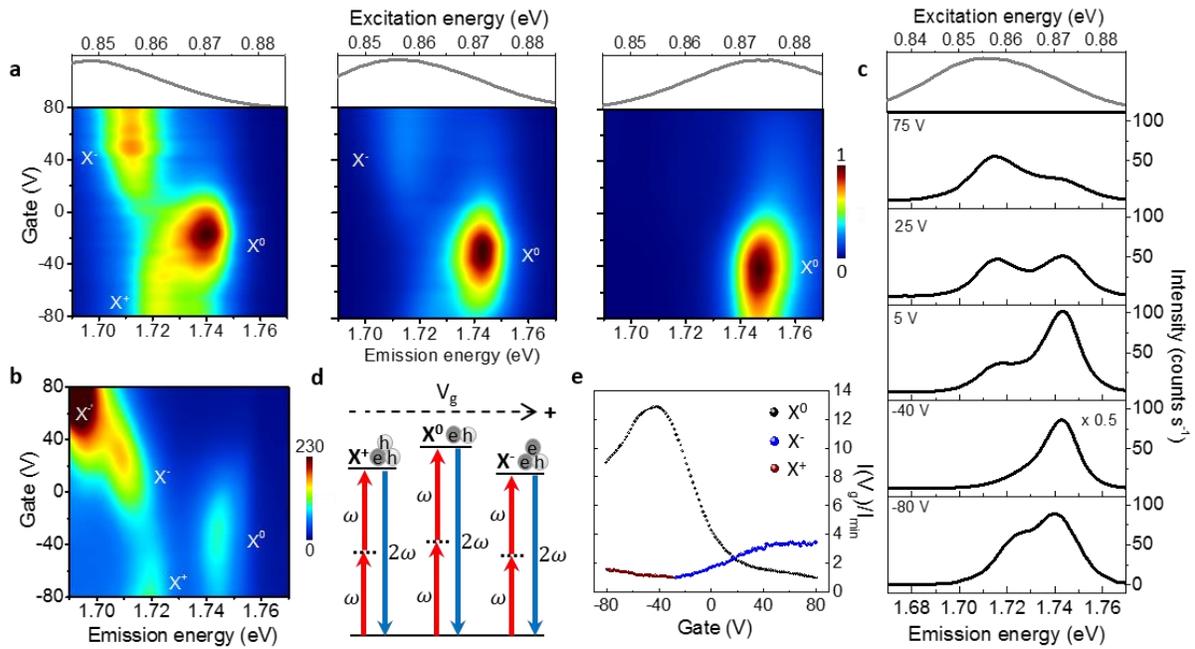

**Figure 3| Electrical control of PL and SHG at 30 K. a**, SHG intensity maps as a function of gate voltage and emission energy. The corresponding excitation laser spectrum is shown on top. From left to right, the plots correspond to two-photon excitation centred on $X^-$, the negative trion $X^-$, and the exciton $X^0$. Each map is normalized individually. **b**, PL intensity (counts s$^{-1}$) map as a function of gate voltage and emission energy showing the excitonic states. **c**, SHG spectra at selected gate voltages. The top plot shows the laser spectrum. **d**, Illustration of the gate-dependent exciton- and trion-enhanced SHG. **e**, Normalized peak intensity of exciton and trion SHG as a function of gate voltage. The laser excitation is fixed at the corresponding resonance.

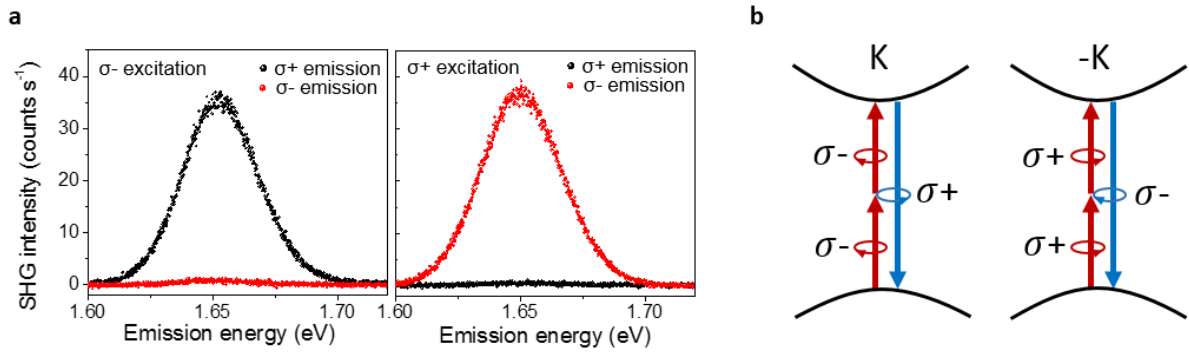

**Figure 4| Valley-dependent SHG selection rules. a**, Circular polarization-resolved SHG spectra showing the generation of counter-circular SHG. **b**, Interband valley optical selection rules for SHG.

# Supplementary Materials:

# Electrical Control of Second-Harmonic Generation in a WSe$_2$ Monolayer Transistor

Kyle L. Seyler, John R. Schaibley, Pu Gong, Pasqual Rivera, Aaron M. Jones, Sanfeng Wu, Jiaqiang Yan, David G. Mandrus, Wang Yao, Xiaodong Xu

**S1. Supplementary Figures**

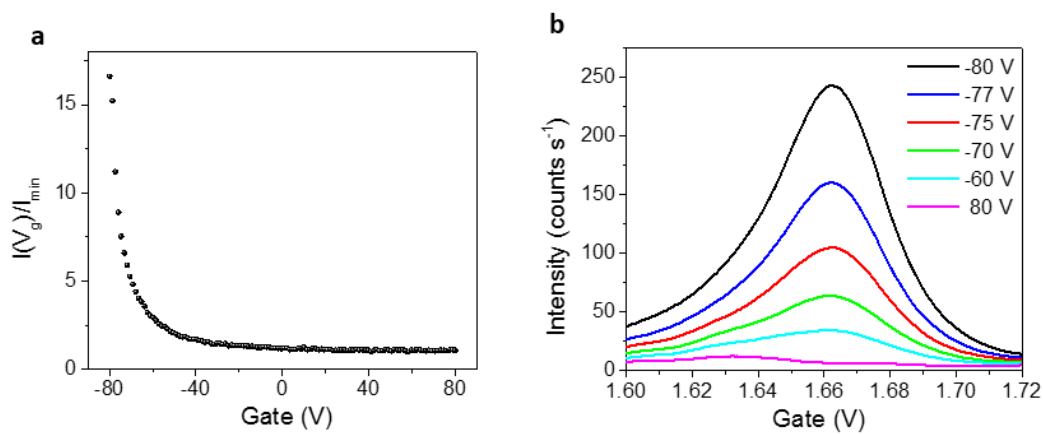

**Supplementary Figure 1| Room-temperature photoluminescence as a function of gate voltage. a**, Normalized peak intensity of PL as a function of gate voltage **b**, PL spectra at selected gate voltages.

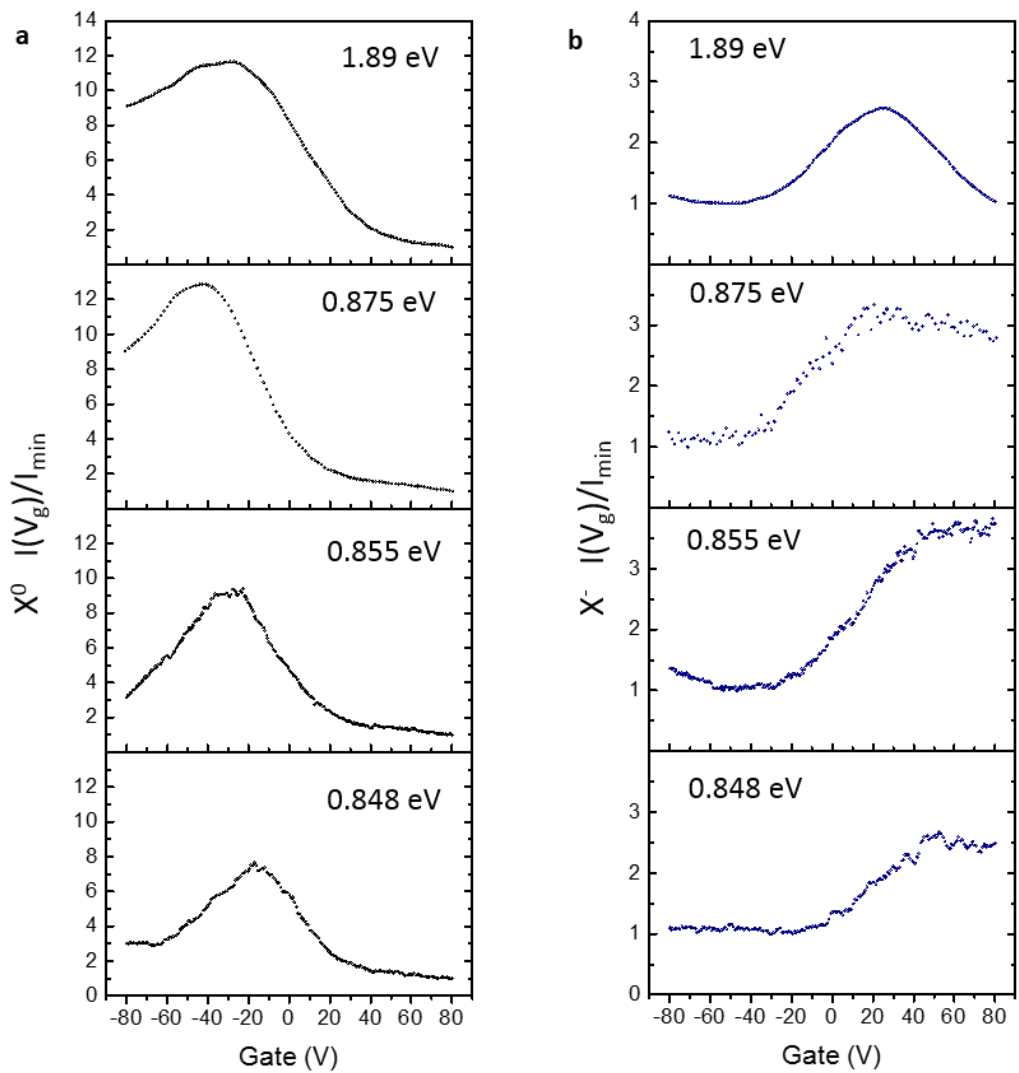

**Supplementary Figure 2| Exciton and trion PL and SHG tunability as a function of gate under different excitation energies. a**, Tunability of exciton PL (top plot) and SHG (bottom 3 plots) as a function of gate voltage. Excitation energies are shown in the corner of each plot. From top to bottom, the three lower plots correspond to two-photon excitation on exciton, trion, and $X^{-'}$. **b**, Tunability of $X^-$ trion PL (top plot) and SHG (bottom 3 plots) as a function of gate voltage. From top to bottom, the three lower plots correspond to two-photon excitation on exciton, trion, and $X^{-'}$.

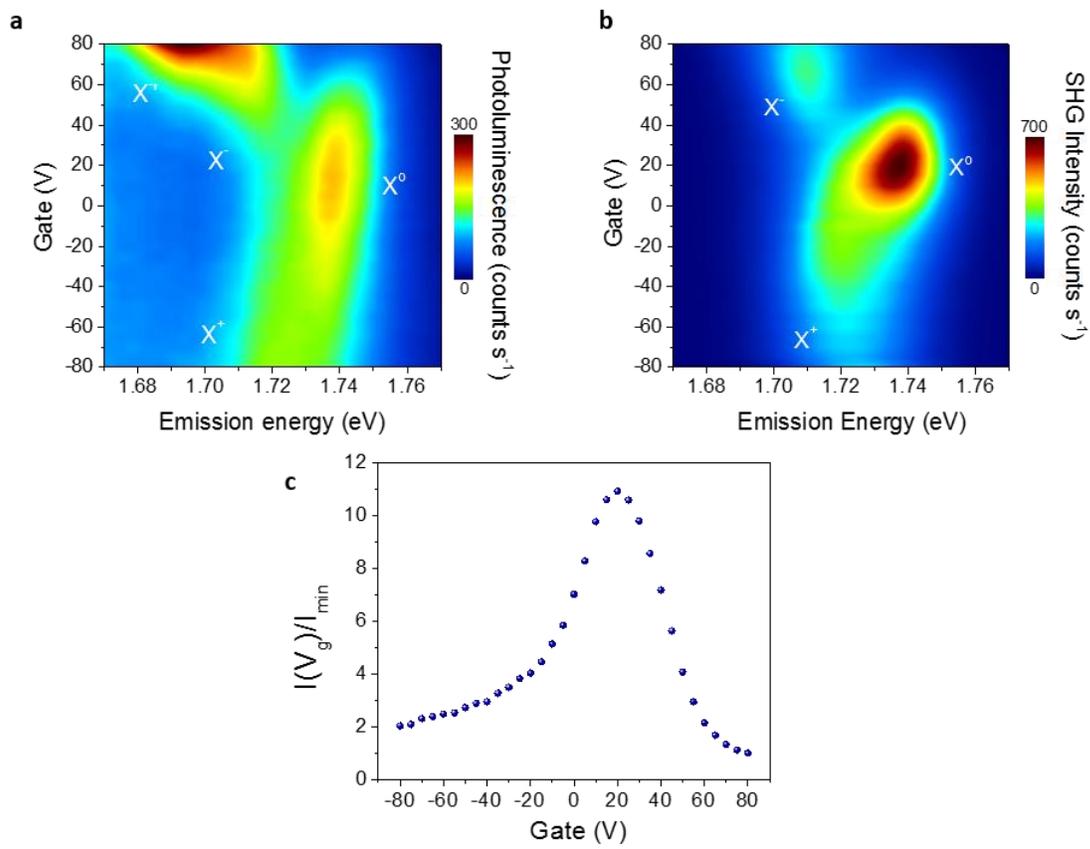

**Supplementary Figure 3| SHG tunability by bias of a second device (30 K). a**, PL intensity map as a function of gate voltage and emission energy, showing the different exciton species. **b**, SHG intensity map showing similar spectral features as Figure 3a. Two-photon excitation energy is centered on $X^-$. **c**, Normalized peak SHG intensity of neutral exciton as a function of gate voltage.

## S2. Microscopic approach to the two-photon selection rule

For monolayer TMDs, one-photon band-edge transitions at the $+K$ $(-K)$ valley are only allowed for $\sigma+$ ($\sigma-$) polarized light[1]. Two-photon interband transitions between the top valence band and lowest conduction band are characterized by the product of matrix elements, $\langle\psi_c|\hat{P}_\pm|\psi_j\rangle\langle\psi_j|\hat{P}_\pm|\psi_v\rangle$, where the $\psi$ give the Bloch states at $\pm K$ for lowest conduction band (c), highest valence band (v), and virtual intermediate states ($j$). $\hat{P}$ is the momentum operator and $\hat{P}_\pm = \hat{P}_x + i\hat{P}_y$. Under the operation of $\hat{C}_3$, a $2\pi/3$ rotation about the $z$-axis, the Bloch states transform as $\hat{C}_3|\psi_j\rangle = e^{-i\frac{2\pi}{3}l_\alpha}|\psi_j\rangle$, where $\alpha = c, v, j$ and $l_\alpha$ is an integer. For rotations about the W atom, $l_\alpha = m_\alpha$, the magnetic quantum number of the local atomic orbitals[2,3]. We can then show how the 3-fold rotational symmetry constrains the two-photon transition matrix elements:

$$\langle\psi_c|\hat{P}_\pm|\psi_i\rangle\langle\psi_i|\hat{P}_\pm|\psi_v\rangle \equiv \langle\psi_c|\hat{C}_3^{-1}\hat{C}_3\hat{P}_\pm\hat{C}_3^{-1}\hat{C}_3|\psi_j\rangle\langle\psi_j|\hat{C}_3^{-1}\hat{C}_3\hat{P}_\pm\hat{C}_3^{-1}\hat{C}_3|\psi_v\rangle$$
$$= \langle\hat{C}_3\psi_c|\hat{C}_3\hat{P}_\pm\hat{C}_3^{-1}|\hat{C}_3\psi_j\rangle\langle\hat{C}_3\psi_j|\hat{C}_3\hat{P}_\pm\hat{C}_3^{-1}|\hat{C}_3\psi_v\rangle$$
$$= e^{i\frac{2\pi}{3}(m_c-m_j\mp 1)}\langle\psi_c|\hat{P}_\pm|\psi_j\rangle e^{i\frac{2\pi}{3}(m_j-m_v\mp 1)}\langle\psi_j|\hat{P}_\pm|\psi_v\rangle,$$

where we used $\hat{C}_3\hat{P}_\pm\hat{C}_3^{-1} = e^{-i\frac{2\pi}{3}}\hat{P}_\pm$. For transitions with two $\sigma+$ or two $\sigma-$ photons, we need $e^{i\frac{2\pi}{3}(m_c-m_v\mp 2)} = 1$, or in other words, $m_c - m_v \mp 2 = 3n$, where $n$ is an integer. At $+K$ $(-K)$, we have $m_c = 0$ and $m_v = +2$ $(-2)$, allowing transitions for two $\sigma-$ $(\sigma+)$ photons. Thus the valley-contrasting optical selection rule extends to two-photon interband transitions, but with a flip in the helicity allowed for each valley. For SHG at $+K$ $(-K)$, two-photon absorption of two $\sigma-$ ($\sigma+$) photons is accompanied by immediate emission of a single $\sigma+$ ($\sigma-$) photon of twice the energy. This leads to the prediction of cross-circularly polarized SHG, fully consistent with the macroscopic approach. Under the $C_3$ rotation, the 1s exciton wavefunction transforms the same as the product of the electron and hole Bloch states at the K points. Thus, the above selection rule is also applicable to the resonant SHG at the 1s exciton.

## Supplementary References